%% file: main.tex
\documentclass[conference]{IEEEtran}
\include{Definition}

\IEEEoverridecommandlockouts
\usepackage{cite}
\usepackage{amsmath,amssymb,amsfonts}
\usepackage{verbatim}
\usepackage{optidef}
\usepackage{textgreek}
\usepackage{algorithmic}
\usepackage{graphicx}
\usepackage[export]{adjustbox}
\usepackage{textcomp}
\usepackage{xcolor}

\newcommand{\mybibliography}{\bibliography{jour_short,conf_short,IEEEexample.bib}}
\DeclarePairedDelimiterXPP\BigOSI[2]%
  {\mathcal{O}}{(}{)}{}%
  {\SI{#1}{#2}}
  
\def\BibTeX{{\rm B\kern-.05em{\sc i\kern-.025em b}\kern-.08em
    T\kern-.1667em\lower.7ex\hbox{E}\kern-.125emX}}
\begin{document}

\title{Deep Unfolding Enabled Constant Modulus Waveform Design for Joint Communications and Sensing\\
}

\author{\IEEEauthorblockN{Prashanth Krishnananthalingam, Nhan Thanh Nguyen, Markku Juntti}
\IEEEauthorblockA{Centre for Wireless Communications, University of Oulu, P.O.Box 4500, FI-90014, Finland}
Emails: \{prashanth.krishnananthalingam, nhan.nguyen, markku.juntti\}@oulu.fi}
%
%
%
%

\maketitle

\begin{abstract}
Joint communications and sensing (JCAS) systems have recently emerged as a promising technology to utilize the scarce spectrum in wireless networks and to reuse the same hardware to save infrastructure costs. In practical JCAS systems, dual functional constant-modulus waveforms can be employed to avoid signal distortion in nonlinear power amplifiers. However, the designs of such waveforms are very challenging due to the nonconvex constant-modulus constraint. The conventional branch-and-bound (BnB) method can achieve optimal solution but at the cost of exponential complexity and long run time. In this paper, we propose an efficient deep unfolding method for the constant-modulus waveform design in a multiuser multiple-input multiple-output (MIMO) JCAS system. The deep unfolding model has a sparsely-connected structure and is trained in an unsupervised fashion. It achieves good communications-sensing performance tradeoff while maintaining low computational complexity and low run time. Specifically, our numerical results show that the proposed deep unfolding scheme achieves a similar achievable rate compared to the conventional BnB method with 30 times faster execution time.
\end{abstract}


\section{Introduction}
The developments in wireless communications systems create the need for more resources, especially for more spectrum which led to spectrum scarcity. This has driven the reuse of spectrum for both communications and radar sensing. The emerging concept of unifying communications and sensing can be referred to by different names, such as integrated sensing and communications (ISAC) \cite{ouyang2022performance, liu2022integrated} or joint communications and sensing (JCAS) \cite{zhang2018multibeam} which is the term used in this paper. 

Early JCAS designs focused on mitigating the mutual interference between the radar and communications subsystems to enable their smooth co-existence \cite{hassanien2016signaling, hassanien2015dual, liu2020joint, ma2020joint, chiriyath2017radar, zheng2019radar, li2016optimum, martone2019joint, liu2018mu, liu2017robust, buzzi2019using}. Specifically, Liu \textit{et al.} \cite{liu2018mu} investigated two multiple-input multiple-output (MIMO) JCAS systems based on sharing the transmit antennas among the communications data and radar probing signal transmission. Hassanien \textit{et al.} \cite{hassanien2015dual} optimized the dual functional waveform by transmitting communications data in the side lobes of the radar waveform, which is a radar-centric design and does not ensure a satisfactory communications performance in general. Wu \textit{et al.} \cite{wu2020waveform} proposed a frequency-hopping MIMO radar-based waveform for channel estimation.

Recent works focus on investigating MIMO JCAS systems to exploit spatial beamforming gains for a good communications--sensing performance tradeoff. The waveform designs were performed with various methods, such as manifold optimization with Riemannian conjugate gradient \cite{8288677}, and semi-definite relaxation \cite{9124713, 9940478}. These works focused on optimizing the transmit waveform under the communications performance constraint. Moreover, the radar Cramer-Rao lower bound \cite{9468975,9420261}, beam pattern matching error \cite{9104378}, and mutual information \cite{9145467} were also considered as the sensing metrics.  In \cite{liu2018toward, tang2022mimo, wu2023constant, liu2021dual}, the practical constant-modulus waveforms were particularly considered and optimized to avoid signal distortion in nonlinear power amplifiers in JCAS systems. Particularly, Liu \textit{et al.} \cite{liu2018toward} investigated the multiuser interference (MUI) minimization problem under the constant-modulus waveform constraint in multiuser MIMO JCAS systems. To tackle the challenging problem, the branch and bound (BnB) method has been employed to achieve optimal performance. However, its complexity is exponential with the number of transmit antennas and the run time is generally long. These challenges motivate us to leverage machine learning (ML), in particular deep learning (DL) techniques for constant-modulus waveform designs.

Machine learning has gained prominence in recent years in various fields of wireless communications \cite{8964474, 8642915, 9772315, 9047156, 9425589, 9626148}. Recently, DL has been leveraged for JCAS designs ~\cite{Mateos2022EndtoEnd,muth2023Autoencoder,xu2022deep,elbir2021terahertz}. Specifically, in \cite{Mateos2022EndtoEnd}, \cite{muth2023Autoencoder}, an end-to-end deep autoencoder model was proposed to learn the joint operations of the beamforming, target detection, and receiver processing. Whereas deep reinforcement learning-based JCAS designs were introduced by Xu \textit{et al.} \cite{xu2022deep} with the focus on sparse transmit arrays. Elbir {\it et al.} considered both model-based and data-based JCAS designs in \cite{elbir2021terahertz}. Specifically, convolutional neural networks and deep neural networks (DNNs) were trained to estimate the direction of the radar targets and for hybrid beamforming \cite{elbir2021terahertz}. Leveraging DL for JCAS enables reliable communications and sensing operations with fixed and low computational complexity and execution time. However, the existing DL-based JCAS designs mostly relied on the learning capabilities of black-box deep neural networks (DNNs). Consequently, they are not explainable as the conventional model-based optimization design methods. Furthermore, their training is generally a challenging task requiring much effort in fine tuning and generating massive data sets. Thus, many conventional optimizers and data-driven DNNs exhibit major limitations due to their resource-constraints, high complexity, and black-box nature. Alternatively, the deep unfolding approach leverage both domain knowledge and learning capabilities to build explainable DL models that achieve performance gains and are easier to implement \cite{9557819}. However, its application for JCAS has not yet been explored.

In this paper, we consider a multiuser MIMO JCAS system. We aim at optimizing the communications-sensing performance tradeoff via minimizing the weighted sum of the MUI and the deviation from the benchmark sensing waveform. In particular, the constant-modulus waveform is considered, as motivated by the work of Liu \textit{et al.} \cite{liu2018toward}. To overcome the challenging design problem, we propose an efficient deep unfolding model based on the projected gradient descent method. The model is trained in an unsupervised strategy to output a reliable constant-modulus waveform. In particular, it has a sparsely connected network architecture, and, thus, it performs the JCAS design with very low computational complexity. 
We finally present extensive simulation results of the proposed scheme in comparison with the conventional BnB approach \cite{8386661}. The simulation results demonstrate that the proposed unfolding network with a few layers and sparse neuron connections performs similarly to the BnB counterpart with significantly lower computational and time complexities.

The remainder of this paper is organized as follows: The system model is described in section \ref{section:2}. Section \ref{section:3} proposes the deep unfolding waveform design and discusses the computational complexity. We provide the numerical results in section \ref{section:4} and conclude the paper in \ref{section:5}. 
\begin{figure}[htbp]
\centerline{\includegraphics[width=0.5\textwidth]{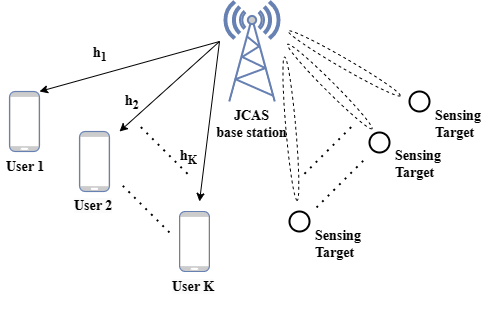}}
\caption{Joint communications and sensing system.}
\label{fig:1}
\end{figure}
\section{System Model}
\label{section:2}
We consider a multiuser MIMO JCAS system, where the base station (BS) equipped with $N$ antennas simultaneously transmits probing signals to the targets and data signals to $K$ 
single-antenna users, as illustrated in Fig. 1. 
Let $\vx_m \in \setC^{N \times 1}$ be the transmit signal vector in the $m$-th communications time frame, and let $\mX = [\vx_1, \ldots, \vx_M]$, with $M$ being the length of the communication frame. Take $\vh_n \in \setC^{K \times 1}$ as the channel vector corresponds to the $n$-th antenna of the BS, and consider $\mH = [\vh_1, \ldots, \vh_N]$. Consider $\vz_m \in \setC^{K \times 1}$ as the noise vector corresponds to the $m$-th communications time frame, where $\vz_m \sim CN(0, N_0\textbf{I}_N )$ and take $\mZ = [\vz_1, \ldots, \vz_M]$.


The received signal at $K$ users can be expressed as
\begin{equation}
\mY = \mH\mX + \mZ.\label{eq:1}
\end{equation}
The transmit signal matrix $ \mX $ is used for both communications and radar sensing. Hence, each communications symbol is also a snapshot of a radar pulse. The downlink channel $ \mH $ is assumed to be flat Rayleigh fading and remains unchanged during a single communications frame/radar pulse. Moreover, the channel is assumed to be perfectly estimated at the BS via pilot symbols.

Let $\vs_m \in \setC^{K \times 1}$ be the desired symbol vector to be received by the users during the $m$-th communications time frame, and let $\textbf{S} = [\vs_1, \ldots, \vs_M]$, whose entries are assumed to be drawn from the same constellation for all users. 
 To unveil the effects of the MUI in the received signals, let us rewrite the signal model (\ref{eq:1}) as
\begin{equation}
\mY = \textbf{S} + (\mH\mX - \textbf{S})  + \mZ,\label{eq:2}
\end{equation}
where the second term, i.e., $(\mH\mX - \textbf{S})$, represents the MUI in the system. The total MUI power can be given as
\begin{equation}
P_{\textrm{MUI}} = \|\mH\mX - \textbf{S}\|_{F}^{2} = \sum_{i,j} |\vh_i^T\vx_j - s_{i,j}|^2, \label{eq:3}
\end{equation}
where $s_{i,j}$ is the (\emph{i, j})-th entry of \textbf{S}, and \begin{math}\mathbb{E}(.)\end{math} represents the ensemble average with respect to time index. The signal-to-interference-plus-noise ratio (SINR) per frame for the $i$-th user is given as
\begin{equation}
\gamma_{i} = \frac{\mathbb{E}(|s_{i,j}|^2)}{\mathbb{E}(|\vh_i^T\vx_j - s_{i,j}|^2) + N_0}. \label{eq:4}
\end{equation}
It is observed that the MUI directly impacts the SINR of the downlink users, which degrades the performance of the communications.

\subsection{Radar Model}

We use the orthogonal chirp waveform matrix, which is adapted from \cite{7450660} as the benchmark waveform, $\mX_0$ for the radar system. Moreover, the transmit beam pattern of the JCAS system is given as



\begin{equation}
P_\textrm{d}(\theta) = \frac{1}{M}\ \textbf{a}^{H}(\theta)\mX\mX^{H}\textbf{a}(\theta), \label{eq:5}
\end{equation}
where, $\theta$ refers to the detection angle and $ \textbf{a}(\theta) = [1, e^{j2\pi\Delta sin(\theta)}, \ldots, e^{j2\pi(N-1)\Delta sin(\theta)}] \in \setC^{N \times 1} $ is the antenna steering vector with $\Delta$ denoting the normalized antenna spacing with respect to the wavelength.

\subsection{Problem Formulation}
We consider a weighted minimization problem which can be formulated below.  Here, the objective function is the weighted sum of the MUI and the distance between $ \mX $ and the desired waveform. 
\begin{mini!}
    {\mX}{\rho\|\mH\mX -\textbf{S}\|_{F}^{2} + (1 - \rho)\|\mX - \mX_0\|_{F}^{2}  \label{eq:7a}}
    {\label{eq:7}}{}
    \addConstraint{|x_{i,j}|}{ = \sqrt{\frac{P_{\textrm{T}}}{N}} \label{eq:7b}},
\end{mini!}
where $ P_{\textrm{T}} $ is the total transmit power, $\mX_0 \in \setC^{N \times 1} $ is the known benchmark radar signal matrix with constant modulus entries, $ \rho \in [0, 1] $ is the weighting factor for the communications systems, and $x_{i,j}$ is the (\emph{i, j})-th entry of $ \mX $. 

The feasible region of the optimization problem (\ref{eq:7}) is a circle of radius $ \sqrt{\frac{P_{\textrm{T}}}{N}} $ for each $ x_{i,j} $ due to the constant-modulus constraint. Thus, the problem is non-convex and NP-hard. 
We next propose a deep unfolding waveform design based on a projected gradient descent approach.

\section{Proposed Deep Unfolding Waveform Design}
\label{section:3}
\subsection{Problem Reformulation}
We herein introduce the general idea to solve problem (\ref{eq:7}) based on the deep unfolding approach. To this end, we first rewrite the two terms in the objective function in (\ref{eq:7a}) as:
\begin{equation}
\|\mH\mX -\textbf{S}\|_{F}^{2} = \sum_{i=1}^M \|\mH\vx_i - \vs_i\|^{2}\label{eq:8}
\end{equation}
and
\begin{equation}
\|\mX - \mX_0\|_{F}^{2} = \sum_{i=1}^M \|\vx_i - \vx_{0i}\|^{2}, \label{eq:9}
\end{equation}
respectively. Here $ \vx_i $, $ \vs_i $ and $ \vx_{0i} $ are the $i$-th columns of $ \mX $, \textbf{S} and $ \mX_0 $ respectively. It is observed from (\ref{eq:8}) and (\ref{eq:9}) that the objective function is column-wise separable. Therefore, we consider an optimization problem which aims to optimize the columns of $ \mX $ separately for the corresponding columns of \textbf{S} and $ \mX_0 $, instead of (\ref{eq:7}). 
Furthermore, as the elements of $ \vx_i $ and $\vx_{0i}$ are constrained to have a constant amplitude, we consider normalized vector variables, $ \vx \in \setC^{N \times 1}$ and $ \vx_0 \in \setC^{N \times 1} $ such that $ \vx_i = \sqrt{\frac{P_{\textrm{T}}}{N}}\vx $ and $  \vx_{0i} = \sqrt{\frac{P_{\textrm{T}}}{N}}\vx_0 $. Here, we have also dropped the column index $ i $ for notational convenience. We reformulate the optimization problem (\ref{eq:7}) using these newly defined normalized vector variables as
\begin{mini!}
{\vx}{\rho \norm{\sqrt{\frac{P_{\textrm{T}}}{N}}\mH\vx-\vs}^{2} + (1 - \rho)\norm{\sqrt{\frac{P_{\textrm{T}}}{N}}(\vx - \vx_{0})}^{2} \label{eq:10a}}
{\label{eq:10}}{}
\addConstraint{|x(n)|}{=1 \label{eq:10b}}, \forall n,
\end{mini!}
where $ x(n) $ denotes the $n$-th entry of $\vx $. 

 Since problem (\ref{eq:7}) and (\ref{eq:10}) are equivalent in the sense that the latter is the column-wise reformulation of (\ref{eq:7}), we focus on (\ref{eq:10}) in the following analysis to develop the deep unfolding network. 

Because neural networks generally work with real-valued data, we present the equivalent real-valued problem of (\ref{eq:10}) as follows:
\begin{mini}
{\boldsymbol{\bar{x}}}{\rho\norm{\sqrt{\frac{P_{\textrm{T}}}{N}}\boldsymbol{\bar{H}\bar{x}}-\boldsymbol{\bar{s}}}^{2} + (1 - \rho)\norm{\sqrt{\frac{P_{\textrm{T}}}{N}}(\boldsymbol{\bar{x}} - \boldsymbol{\bar{x}}_{0})}^{2},}
{}{} \label{eq:11}
\end{mini}
where, $ \boldsymbol{\bar{s}} = 
  \begin{bmatrix}
    \Re(\vs)\\ 
    \Im(\vs)
  \end{bmatrix} \in \mathbb{R}\ ^{{2K}},
\boldsymbol{\bar{x}} = 
  \begin{bmatrix}
    \Re(\vx)\\ 
    \Im(\vx)
  \end{bmatrix} \in \mathbb{R}\ ^{{2N}},
\boldsymbol{\bar{x}}_{0} = 
  \begin{bmatrix}
    \Re(\vx_{0})\\ 
    \Im(\vx_{0})
  \end{bmatrix} \in \mathbb{R}\ ^{{2N}},  and
\boldsymbol{\bar{H}} = 
  \begin{bmatrix}
    \Re(\mH) & -\Im(\mH)\\ 
    \Im(\mH) & \Re(\mH)
  \end{bmatrix} \in \mathbb{R}\ ^{{2K}\times{2N}} $

In light of the projected gradient-descent method, the iterative update procedure of the waveform in the $p$-th iteration can be expressed as
\begin{equation}
\boldsymbol{\hat{x}}_{p} =\prod \left(\boldsymbol{\hat{x}}_{p-1} - \delta_{p-1} \frac{\partial f(\boldsymbol{\bar{x}})}{\partial \boldsymbol{\bar{x}}} \vline_{\boldsymbol{\bar{x}} = \boldsymbol{\hat{x}}_{p-1}}\right), \label{eq:12} 
\end{equation}   
where $\boldsymbol{\hat{x}}_{p}$ denotes the output vector of the neural network in the $p$-th iteration, $ \delta_{p-1} $ denotes the step size, and $ \Pi (\cdot) $ denotes the nonlinear projection operator and  $f(\cdot)$ is the objective function in (\ref{eq:11}), i.e., $ \rho\norm{\sqrt{\frac{P_{\textrm{T}}}{N}}\boldsymbol{\bar{H}\bar{x}}-\boldsymbol{\bar{s}}}^{2} + (1 - \rho)\norm{\sqrt{\frac{P_{\textrm{T}}}{N}}(\boldsymbol{\bar{x}} - \boldsymbol{\bar{x}}_{0})}^{2} $. 
The iterative update procedure in (\ref{eq:12}) motivates the application of a DNN of $ L $ layers to output the transmit waveform. Specifically, we aim at developing a deep unfolding DNN architecture mimicking the nonlinear transformation in (\ref{eq:12}). As long as the DNN is well constructed and trained, it can generate the solution vector $ \boldsymbol{\bar{x}} $ of the problem (\ref{eq:11}) for the given weight factor $ \rho $. 
We detail the structure and training of the DNN next.
\subsection{Deep Unfolding Model's Structure and Training}
\begin{figure}[htbp]
\centerline{\includegraphics[width=0.5\textwidth]{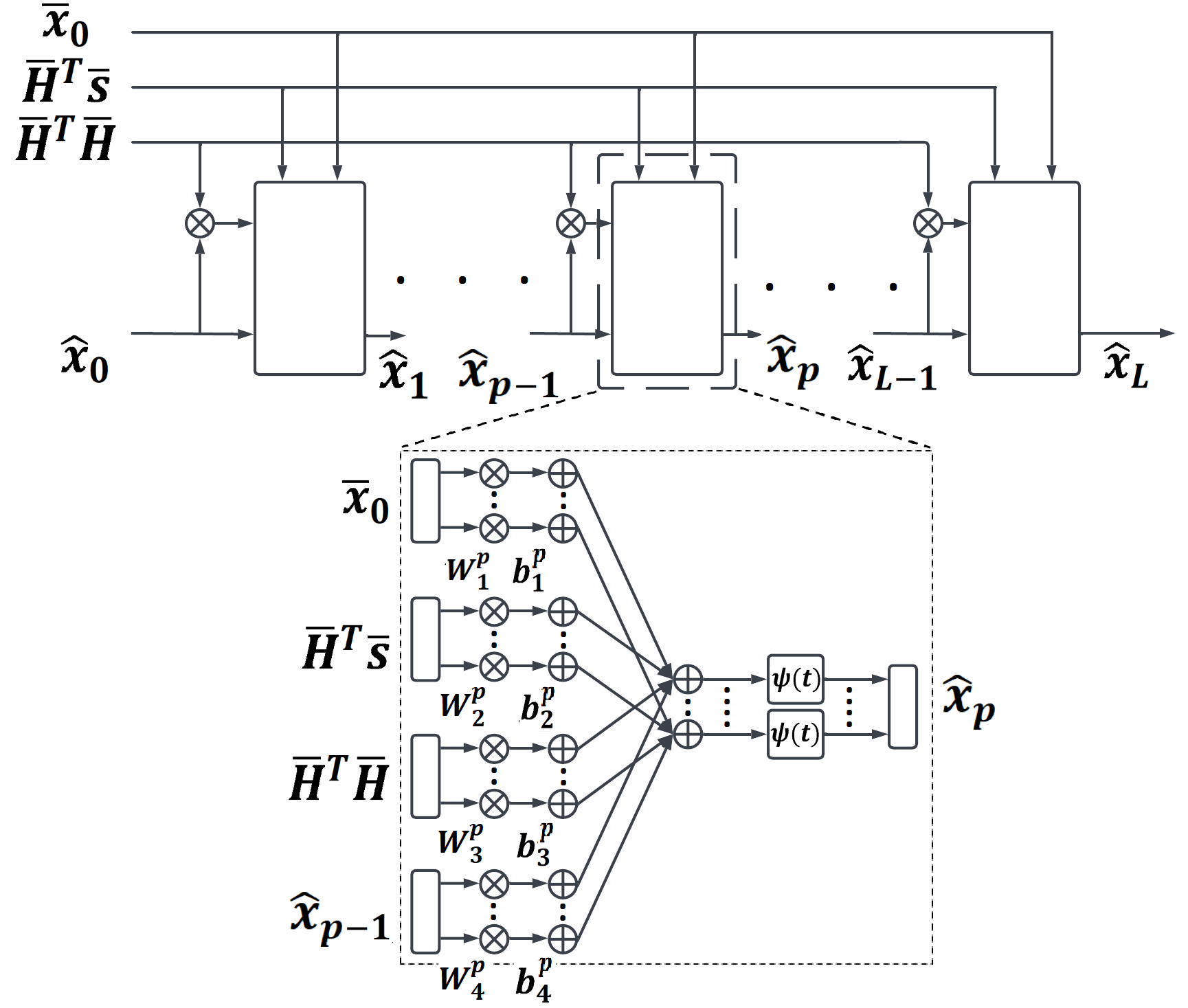}}
\caption{Architecture of the neural network.}
\label{fig:2}
\end{figure}

From (\ref{eq:11}) and (\ref{eq:12}) we have
\begin{multline}
\boldsymbol{\hat{x}}_{p} =\prod \Bigg( \Bigg.\boldsymbol{\hat{x}}_{p-1} - \delta_{p-1} \Bigg( \Bigg.2\rho \frac{P_{\textrm{T}}}{N} \boldsymbol{\bar{H}}^T\boldsymbol{\bar{H}}\boldsymbol{\hat{x}}_{p-1} - 2\rho \sqrt{\frac{P_{\textrm{T}}}{N}} \boldsymbol{\bar{H}}^T\boldsymbol{\bar{s}}\\ + 2(1-\rho) \frac{P_{\textrm{T}}}{N} \boldsymbol{\hat{x}}_{p-1} - 2(1-\rho) \frac{P_{\textrm{T}}}{N} \boldsymbol{\bar{x}}_{\textrm{0}}\Bigg. \Bigg) \Bigg. \Bigg). \label{eq:14}
\end{multline}
%
%
Here, $\boldsymbol{\hat{x}}_{p}$ is the estimated transmit signal vector in the $p$-th iteration. Equation (\ref{eq:14}) shows that it depends on $\boldsymbol{\bar{x}_{0}}$, $\boldsymbol{\bar{H}}^T\boldsymbol{\bar{s}}$, $\boldsymbol{\bar{H}}^T\boldsymbol{\bar{H}}$ and $\boldsymbol{\hat{x}}_{p-1}$. Therefore, these variables are used as inputs to the deep unfolding network. Fig. \ref{fig:2} shows the architecture of the proposed neural network. There are four vector inputs in each layer, namely, $\boldsymbol{\bar{x}_{0}}$, $\boldsymbol{\bar{H}}^T\boldsymbol{\bar{s}}$, $\boldsymbol{\bar{H}}^T\boldsymbol{\bar{H}}$ and $\boldsymbol{\hat{x}}_{p-1}$ 
and the vector output of the corresponding layer is $\boldsymbol{\hat{x}}_{p}$, the estimation of the unknown transmitted vector. The weight matrices for each input are denoted by $\boldsymbol{W}_{1}^{p}$, $\boldsymbol{W}_{2}^{p}$, $\boldsymbol{W}_{3}^{p}$ and $\boldsymbol{W}_{4}^{p}$. The bias vectors for each input are denoted by $\boldsymbol{b}_{1}^{p}$, $\boldsymbol{b}_{2}^{p}$, $\boldsymbol{b}_{3}^{p}$ and $\boldsymbol{b}_{4}^{p}$. By inspecting (\ref{eq:14}), only the same indexed elements of the input and output vectors are related. Therefore, the neural network has a sparsely connected structure, where the input and the output are connected element-wise in each layer. This helps to reduce the computational complexity significantly. Due to the sparse connections, the weight matrices become diagonal matrices. The activation function is  denoted by $\psi(t)$, which is defined as
\begin{equation}
\psi(t) = -1 + 2(\sigma(t+0.5)-\sigma(t-0.5)),
\end{equation}
where $\sigma(t)$ denotes the rectified linear unit (ReLU) function.
The output of the $p$-th layer is updated as follows:

\begin{equation}
\boldsymbol{m}_{1}^{p} = \boldsymbol{W}_{1}^{p}\boldsymbol{\bar{x}}_{0} + \boldsymbol{b}_{1}^{p}, \label{eq:15}
\end{equation}
\begin{equation}
\boldsymbol{m}_{2}^{p} = \boldsymbol{W}_{2}^{p}\boldsymbol{\bar{H}}^T\boldsymbol{\bar{s}} + \boldsymbol{b}_{2}^{p}, \label{eq:16}
\end{equation}
\begin{equation}
\boldsymbol{m}_{3}^{p} = \boldsymbol{W}_{3}^{p}\boldsymbol{\bar{H}}^T\boldsymbol{\bar{H}\hat{x}}_{p-1} + \boldsymbol{b}_{3}^{p}, \label{eq:17}
\end{equation}
\begin{equation}
\boldsymbol{m}_{4}^{p} = \boldsymbol{W}_{4}^{p}\boldsymbol{\hat{x}}_{p-1} + \boldsymbol{b}_{4}^{p}, \label{eq:18}
\end{equation}
\begin{equation}
\boldsymbol{s}_p = \boldsymbol{m}_{1}^{p} + \boldsymbol{m}_{2}^{p} + \boldsymbol{m}_{3}^{p} + \boldsymbol{m}_{4}^{p}, \label{eq:19}
\end{equation}
\begin{equation}
\boldsymbol{\hat{x}}_p = \psi(\boldsymbol{s}_p). \label{eq:20}
\end{equation}

To avoid the vanishing gradients, the sensitivity to initialization, and the saturation of activation functions, we leverage a loss function that covers the output from all the layers. This technique is inspired by the idea of auxiliary classifiers discussed in GoogLeNet \cite{7298594}. The loss function can be given as,
\begin{multline}
l( \boldsymbol{x}_{opt} ; \boldsymbol{\hat{x}} \boldsymbol{(\bar{H},\bar{s})}) = \sum_{p=1}^L \Bigg( \Bigg.\rho\norm{\sqrt{\frac{P_T}{N}}\boldsymbol{\bar{H}}\boldsymbol{\hat{x}}_{p}-\boldsymbol{\bar{s}}}^{2} \\ + (1 - \rho)\norm{\sqrt{\frac{P_T}{N}}(\boldsymbol{\hat{x}}_{p} - \boldsymbol{\bar{x}}_{0})}^{2}\Bigg. \Bigg).\label{eq:13}  
\end{multline}
\subsection{Computational Complexity}
\label{subsec:3a}
We first note that $\boldsymbol{\bar{H}}^T\boldsymbol{\bar{s}}$ and $\boldsymbol{\bar{H}}^T\boldsymbol{\bar{H}}$ are computed beforehand and fed to the network.
Since $\boldsymbol{\bar{H}}^T\boldsymbol{\bar{H}} \in $ \begin{math}\mathbb{R}\ ^{{2N}\times{2N}}\end{math}, computing $ \boldsymbol{\bar{H}}^T\boldsymbol{\bar{H}}\boldsymbol{\hat{x}}_{p-1} $ requires a total of $ 2N(4N-1) $ floating point operations (FLOPs). The multiplications in (\ref{eq:15}) - (\ref{eq:18})   need  $ 8N $ FLOPs. Furthermore, additions in (\ref{eq:15}) - (\ref{eq:18})   need  $ 8N $ FLOPs. The additions in (\ref{eq:19}) need $ 6N $ FLOPs. The total arithmetic operations per layer will be $ 2N(4N+11) $ FLOPs. Hence, the computational complexity is $\mathcal{O} (2N(4N+11)L)$. As the computational complexity of the BnB is exponential with $ N $, it is higher compared to the proposed deep unfolding scheme. \cite{Tuy2016}

\section{Numerical Results}
\label{section:4}
In this section, we present the numerical results to validate the proposed deep unfolding network. 
The total power budget is set to $ P_{\textrm{T}} = 30$ $ dBm $ and the channel is assumed to be flat Rayleigh fading. Moreover, the losses at the transmitter and receiver are assumed to be cancelled by the amplifier gains at the transmitter and receiver respectively. 
Each entry of the channel matrix $\mH$ is assumed to follow the standard complex Gaussian distribution. 
Moreover, we set  the number of transmit antennas as $ N $ = 8 and assume that the ULA with half-wavelength spacing is deployed at the BS. The desired symbol matrix \textbf{S} is assumed to be populated with symbols from the unit-power QPSK alphabet. 
The length of the communication frame/radar pulse is set as $ M $ = 20. The number of layers for the neural network is set as $L$ = 10. 
We train the network with batches of 100 channel realizations. For  the benchmark radar signal, a constant modulus chirp signal is used. The noise signal is assumed to be complex additive white Gaussian (AWG) with zero mean and the variance of $N_0$. The transmit SNR is defined as SNR = \begin{math}P_{\textrm{T}}/N_0\end{math}.

The neural network is implemented in Python with the Tensorflow library. An initial learning rate of 0.0001 and a decaying factor of 0.97 are used. Adam optimizer is used to optimize the neural network. The input $\boldsymbol{\hat{x}}_{\textrm{0}}$ is initialized as a vector of zeros.

\begin{figure}[htbp]
\centerline{\includegraphics[width=0.5\textwidth]{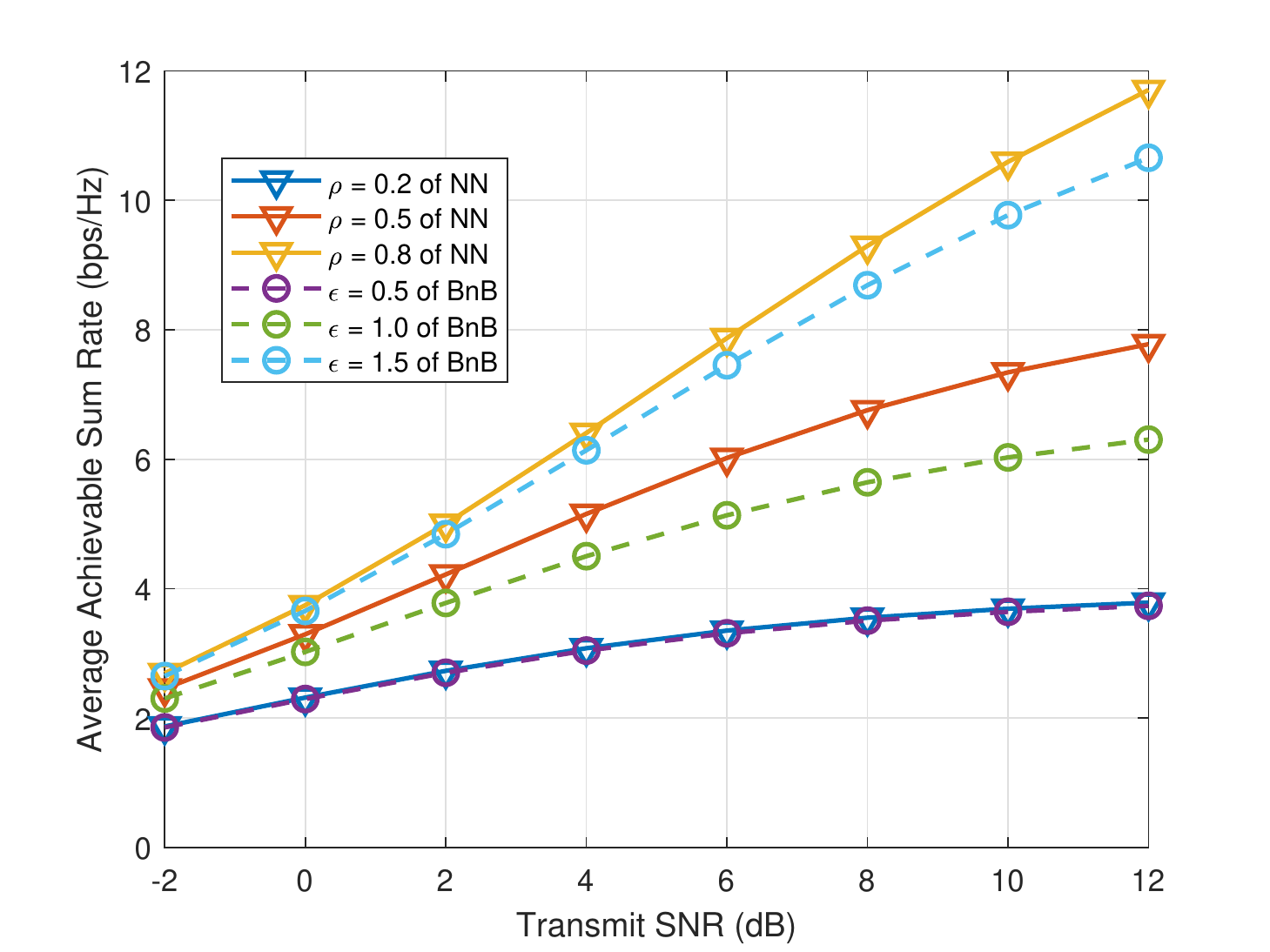}}
\caption{Average achievable sum rate with SNR for different approaches.}
\label{fig:3}
\end{figure}

Fig. \ref{fig:3} shows the system average achievable sum rate with SNR for different \(\rho\) and \(\epsilon\). Here, \(\epsilon\) corresponds to normalized tolerable difference defined in \cite{8386661}. 

\begin{figure}[htbp]
\centerline{\includegraphics[width=0.5\textwidth]{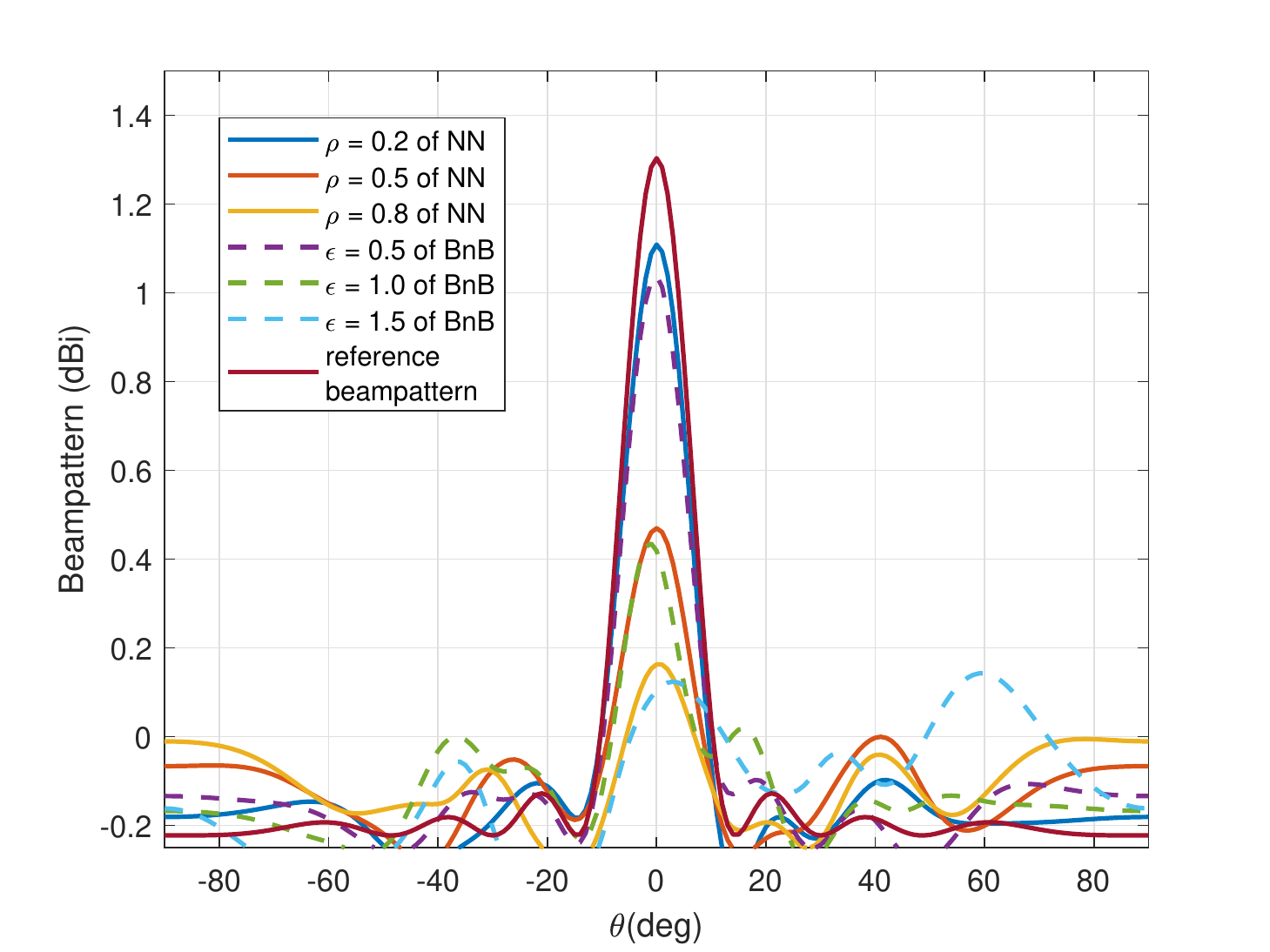}}
\caption{Radar beam patterns obtained by different approaches.}
\label{fig:4}
\end{figure}

Fig. \ref{fig:4} shows the system average radar beam pattern for different \(\rho\) and \(\epsilon\). The reference beam pattern is also included for the comparison. 
It can be observed from Fig. \ref{fig:4} that the designed beam pattern through the neural network tends to match with the reference waveform when \(\rho\) is lower. Moreover, by comparing the results of neural network and BnB, it can be stated that the neural network keeps the main lobe in the desired direction even though the transmit power gets reduced while the beam pattern gets a significant degradation for BnB. Furthermore, by observing the sum rate and beam pattern results for \(\rho\) = 0.8 and \(\epsilon\) = 1.5, we could state that the neural network maintain both a better beam pattern and better sum rate response compared to BnB.

\begin{figure}[htbp]
\centerline{\includegraphics[width=0.5\textwidth]{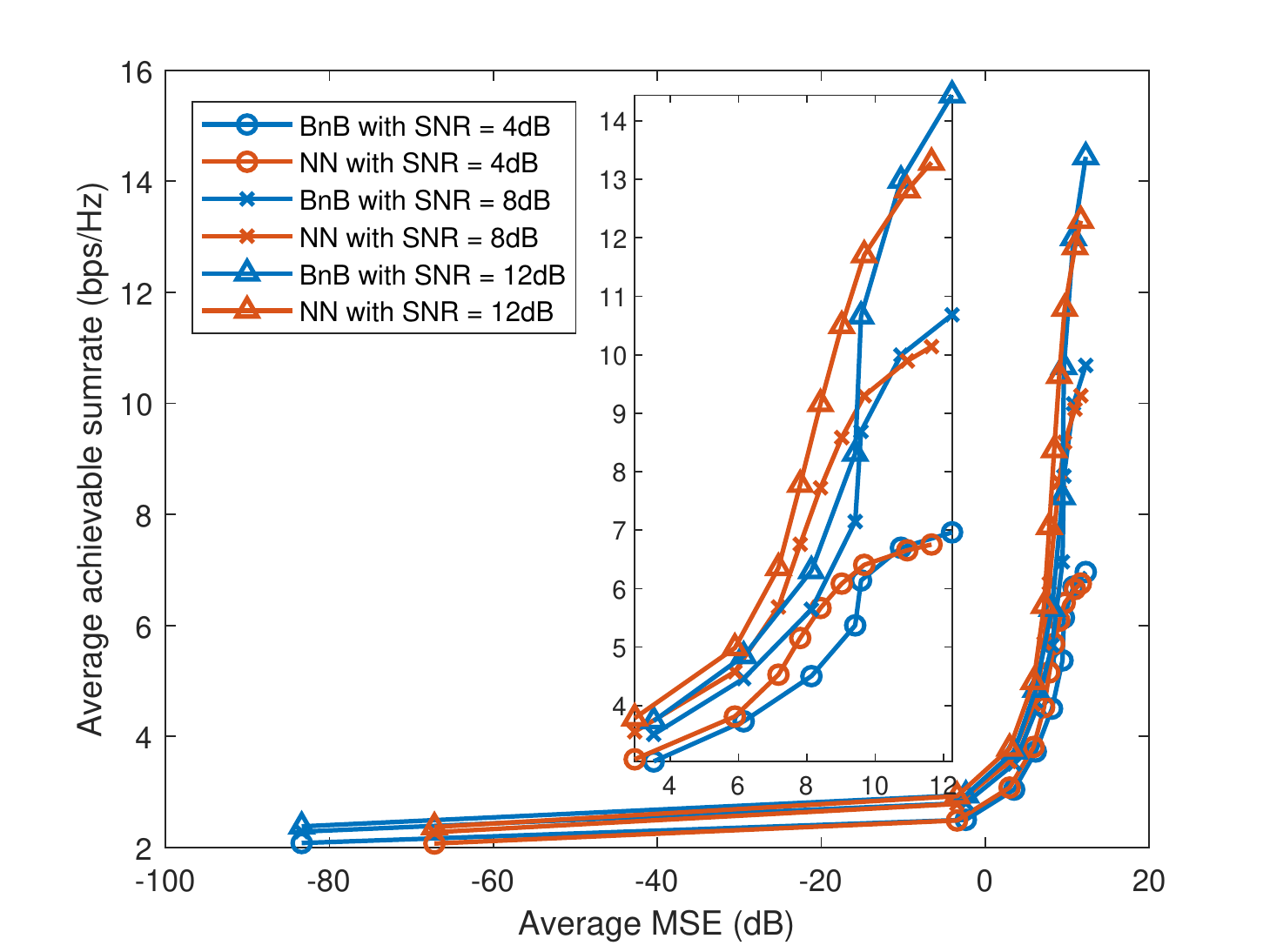}}
\caption{Variation of average achievable sum rate with average MSE of the radar beam pattern.}
\label{fig:5}
\end{figure}

We use the results for the constant modulus waveform design in \cite{8386661} through the BnB algorithm as a benchmark to validate our results. 
We compare the system average sum rate with the average radar mean squared error (MSE) for both methods. Fig. \ref{fig:5} shows the variation of sum rate with average MSE of the radar beam pattern for different SNR values. The variation in Fig. \ref{fig:5} implies that the proposed neural network performs almost the same as the BnB algorithm. Moreover, a closer inspection of the results shows that the results of neural network are slightly better in most cases.

We analyze the time complexities of both methods next. We compare their run times in the same machine. The BnB algorithm was implemented and tested through MATLAB. We kept all the values of the variables except $N$ as same as the values mentioned at the beginning of the section. Table \ref{tab:1} shows the average run time per channel for both methods with different transmit antennas. We consider the communication-only case for this purpose. The per-channel execution times for the neural network are significantly lower compared to the BnB. Based on the results, it can be stated that the time complexity of the neural network is lower than BnB.

\begin{table}[htbp]
\caption{Run times for both BnB and NN methods with number of transmit antennas}
\begin{center}
\begin{tabular}{|c|c|c|}
\hline
\textbf{Transmit}&\multicolumn{2}{|c|}{\textbf{Algorithm}} \\
\cline{2-3} 
\textbf{Antennas} & \textbf{\textit{Branch and Bound}}& \textbf{\textit{Neural Network}} \\
\hline
8 & 15.9 s & 0.507 s  \\
\hline
16 & 17.4 s & 0.532 s  \\
\hline
\end{tabular}
\label{tab:1}
\end{center}
\end{table}

\section{Conclusions}
\label{section:5}
In this paper, we proposed a neural network-based solution to the problem of waveform optimization in JCAS systems. Considering the results in section \ref{section:4} along with the complexity analysis in section \ref{subsec:3a}, it can be stated that our scheme gives almost the same performance as the BnB algorithm while experiencing reductions in both computational and time complexities. 
Simulation results validated the claim that our neural network performs as same as the BnB.

\section*{Acknowledgement}
This research has been supported by the Academy of Finland, 6G Flagship program and Infotech Oulu under Grant 346208.

\bibliographystyle{IEEEtran}
\mybibliography

\vspace{12pt}

\end{document}

%% file: Definition.tex
\usepackage{amsmath,graphicx}
\usepackage{color}
\usepackage{graphicx}
\usepackage{epstopdf}
\usepackage{amsmath}
\usepackage{amssymb}
\usepackage{mathrsfs}
\usepackage{hyperref}
\usepackage{tikz}
\usepackage{bm}
\usepackage[english]{babel}
\usepackage{cite}
\usepackage{rotfloat}
\usepackage{mathtools}
\usepackage[font=normalsize,labelfont=bf]{caption}
\usepackage{empheq}

\usepackage{amsmath}
\usepackage{makecell}
\usepackage{algorithm,algorithmic}
\usepackage{multirow}
\usepackage{subfigure}
\usepackage{booktabs}
\usepackage{colortbl}
\usepackage{multirow}
\usepackage{hhline}
\usepackage{stfloats}
\usepackage{multicol}
\usepackage{bbm}
\usepackage{cases}
\graphicspath{ {Figures/} }
\newsavebox{\foobox}

\definecolor{kugray5}{RGB}{224,224,224}

\usepackage[normalem]{ulem}
\newcommand\rsout{\bgroup\markoverwith
	{\textcolor{red}{\rule[0.5ex]{2pt}{0.8pt}}}\ULon}



\makeatletter
\newcommand{\ALOOP}[1]{\ALC@it\algorithmicloop\ #1%
	\begin{ALC@loop}}
	\newcommand{\ENDALOOP}{\end{ALC@loop}\ALC@it\algorithmicendloop}

\makeatother

\usepackage{etoolbox}
\let\mybibitem\bibitem
\renewcommand{\bibitem}[1]{%
	\ifstrequal{#1}{nature}
	{\color{blue}\mybibitem{#1}}
	{\color{black}\mybibitem{#1}}%
}

\graphicspath{ {Figures/} }

\DeclareCaptionLabelSeparator{periodspace}{.\quad}

\renewcommand{\figurename}{Fig.}
\addto\captionsenglish{\renewcommand{\figurename}{Fig.}}
\newcommand{\norm}[1]{\left\lVert#1\right\rVert} 

\allowdisplaybreaks

\usepackage{setspace}

\newcommand{\mH}{{\mathbf{H}}}

\newcommand{\mX}{{\mathbf{X}}}
\newcommand{\mY}{{\mathbf{Y}}}

\newcommand{\mZ}{{\mathbf{Z}}}

\newcommand{\setC}{\mathbb{C}} 

\newcommand{\vs}{{\mathbf{s}}}
\newcommand{\vx}{{\mathbf{x}}}

\newcommand{\vz}{{\mathbf{z}}} 
\newcommand{\vh}{{\mathbf{h}}}

\def\Re{\mathop{\mathrm{Re}}}
\def\Im{\mathop{\mathrm{Im}}}

\def\b0{{\pmb{0}}}